\documentclass[sigconf, nonacm]{acmart}

\usepackage{algorithm}
\usepackage[noend]{algorithmic}
\usepackage{enumitem}
\usepackage{soul, color, xcolor}
\usepackage[table, dvipsnames]{xcolor}
\usepackage{multirow} 
\usepackage[table,xcdraw]{xcolor} 
\usepackage{xcolor}

\AtBeginDocument{%
  }

\setcopyright{none} 

\begin{document}

\title{A Spatio-Temporal Expert Prefetching Framework for Efficient MoE-based LLM Inference}

\author{Yingnan Zhao}
\affiliation{
  \institution{George Washington University}
  \city{Washington}
  \state{DC}
  \country{USA}
}
\email{yzhao96@gwu.edu}

\author{Razvan Bunescu}
\affiliation{
  \institution{University of North Carolina at Charlotte}
  \city{Charlotte}
  \state{NC}
  \country{USA}
}
\email{razvan.bunescu@charlotte.edu}

\author{Ahmed Louri}
\affiliation{
  \institution{George Washington University}
  \city{Washington}
  \state{DC}
  \country{USA}
}
\email{louri@gwu.edu}

\author{Avinash Karanth}
\affiliation{
  \institution{Ohio University}
  \city{Athens}
  \state{OH}
  \country{USA}
}
\email{karanth@ohio.edu}

\author{Ke Wang}
\affiliation{
  \institution{University of North Carolina at Charlotte}
  \city{Charlotte}
  \state{NC}
  \country{USA}
}
\email{ke.wang@charlotte.edu}


\begin{abstract}

Mixture-of-Experts (MoE) based large language models (LLMs), such as Qwen and DeepSeek, have recently emerged as an effective approach to improving model capacity without proportionally increasing computational cost. By replacing the conventional feed-forward network in dense LLMs with a set of experts and activating only a subset of them for each input token, MoE models significantly increase the total number of parameters while keeping the per-token computation relatively manageable. However, this dynamic and irregular expert activation pattern also introduces substantial expert loading overhead during inference, since the required experts must be fetched on demand according to token-dependent routing results. As a result, expert loading latency becomes a major source of performance and energy inefficiency. Existing approaches attempt to address this challenge through methodologies such as adaptive routing and pre-gating. These methods often suffer from inference accuracy degradation or introduce additional training overhead, limiting their practical benefits.

To this end, we first perform a comprehensive analysis of expert selection behavior in various MoE-based LLMs and applications, including language understanding and code generation. Our analysis reveals that, within each application domain, expert requests exhibit strong correlation across both adjacent MoE layers and consecutive decoding tokens, making future expert activations predictable. Based on this insight, we propose ST-MoE, a spatio-temporal expert prefetching framework that proactively stages experts ahead of use to overlap expert loading with ongoing computation. ST-MoE combines a lightweight runtime prediction mechanism that preserves the original routing behavior with a reconfigurable hardware design that efficiently supports dynamic expert prefetching. The combined effect of the prediction mechanism with the supporting hardware significantly improves MoE inference performance and energy efficiency while preserving model inference accuracy. Evaluation results on real-world applications show that ST-MoE achieves 85\% expert prediction accuracy and delivers average speedups of 2.5x, 2.2x, and 1.5x, as well as average energy efficiency improvements of 2.5x, 1.8x, and 2.0x, compared to GPU, Adap-Gating, and Pre-gated MoE, respectively.

\end{abstract}

\maketitle

\section{Introduction}
\label{sec: intro}

Large Language Models (LLMs) have achieved remarkable success across a wide range of applications, such as strategic reasoning~\cite{zhang2024scaling,zhang2024llm}, language understanding~\cite{kim2024understanding,nam2024using}, code generation~\cite{fakhoury2024llm,huang2025bias}, and multimodal intelligence~\cite{hu2024bliva,zhan2024anygpt,wu2024next}. A key driver of this progress has been the continued scaling of model capacity, which has consistently improved LLM capability across diverse applications. However, scaling dense LLMs often incurs rapid increase in computation and memory costs, making it increasingly challenging for efficient deployment. To address this issue, the Mixture-of-Experts (MoE) technique is now widely used in both frontier and open-source LLMs ~\cite{bi2024deepseek,yang2025qwen3,jiang2024mixtral}. By replacing the conventional dense feed-forward network with a set of experts and activating only a subset of them for each input token, MoE-based LLMs increase the total number of parameters while significantly decreasing the number of parameters that are actively used per token. However, the decrease in the number of parameters comes at the cost of a more complex computational path during inference.

For each token, the gating function dynamically selects only a small subset of experts, making the computation path highly input dependent. The experts required by one token may differ from those required by the next token or by the next MoE layer, making expert requests difficult to schedule statically. Since each expert contains a large number of parameters, keeping all experts on-chip is impractical due to limited buffer capacity. Instead, the required expert weights must be fetched on demand from off-chip memory, which incurs substantial data movement and memory access latency. This challenge becomes even more critical during the decoding phase when tokens are generated sequentially. Consequently, any delay in loading the required experts can stall subsequent computation, reduce hardware utilization, and increase energy consumption. Therefore, although MoE lowers the computational cost for LLMs, inefficient expert loading can significantly limit the performance and energy efficiency benefits.

To address the inference challenges of MoE-based LLMs, prior work has explored several approaches, including adaptive routing, expert prefetching, and expert caching. For example, Adap-gating~\cite{li2023adaptive} improves inference efficiency by dynamically reducing the number of activated experts, while Pre-gated MoE~\cite{hwang2024pre} introduces an additional trained pre-gating function together with CPU-GPU co-design to proactively transfer the experts required by the next MoE block. Despite their advantages, these approaches still face multiple limitations for effective deployment. Adap-gating alters the routing behavior learned under the original training policy, which can lead to significant inference accuracy degradation, whereas Pre-gated MoE depends on an additional trained component, which imposes extra training overhead and deployment complexity. More broadly, existing methods fail to fully exploit structured expert activation patterns that naturally emerge during decoding. As a result, they leave significant room to improve MoE inference efficiency in a practical, adaptive, and accuracy-preserving manner.

\begin{figure}
    \centering
    \includegraphics[width=\linewidth]{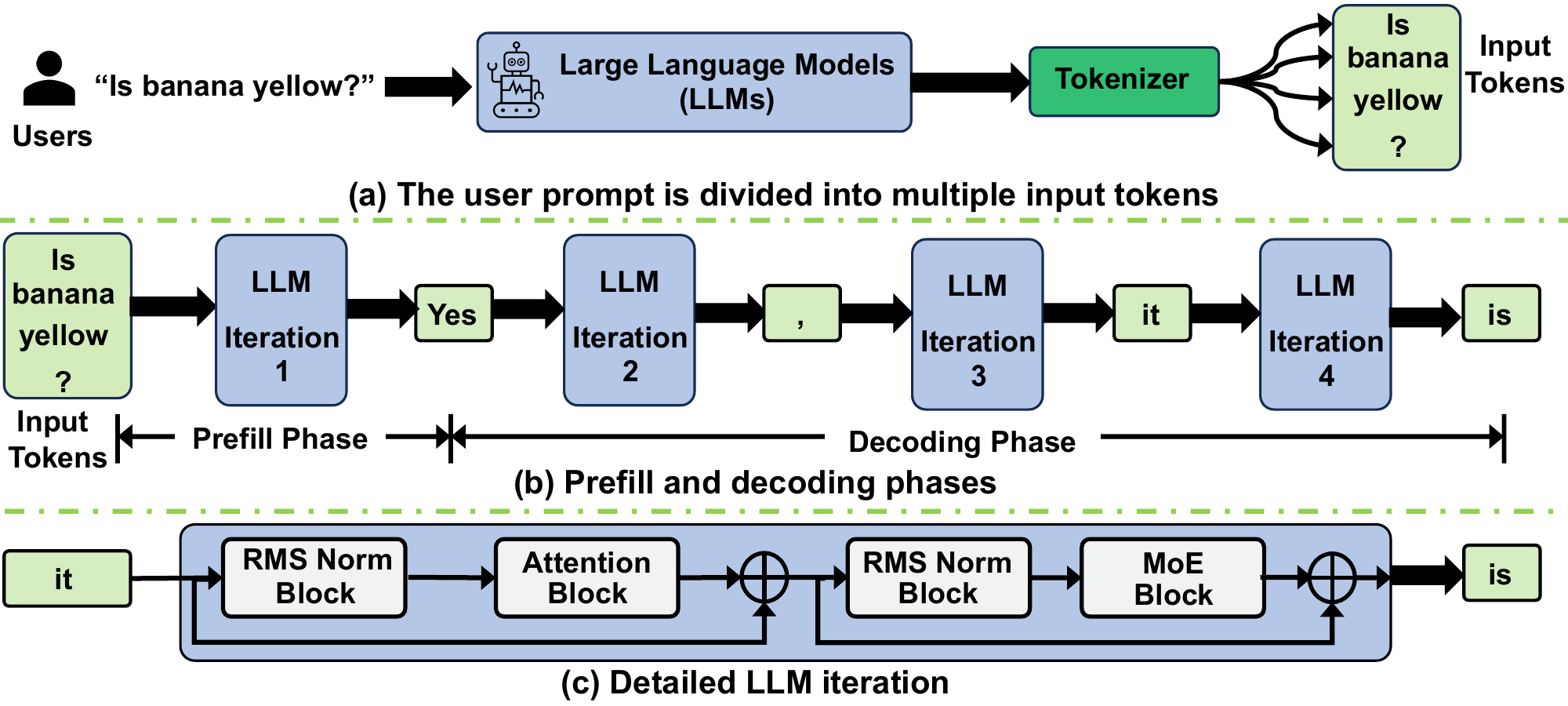}
    \caption{(a) The user prompt is tokenized into multiple input tokens by the LLM tokenizer. (b) LLM inference consists of two phases: prefill and decoding. (c) One LLM iteration consists of three types of blocks: Root Mean Square (RMS) Normalization, attention, and Mixture-of-Experts (MoE). DeepSeek V2 is used here as a representative LLM model.}
    \label{fig: figure 1}
\end{figure}

To this end, we conduct a comprehensive empirical analysis of expert selection behavior in representative MoE-based LLMs across diverse applications. Our study reveals two important types of correlations in the MoE mechanism when used with LLMs: expert selections are strongly correlated across adjacent MoE layers (spatial) and between consecutive decoded tokens (temporal), making expert requests amenable to predictability at runtime. Building on this critical insight, we propose ST-MoE, a spatial-temporal expert prefetching framework that proactively stages experts before they are needed to overlap expert loading latency with ongoing computation. ST-MoE integrates a lightweight dynamic prediction mechanism with a reconfigurable hardware architecture to enable efficient expert prefetching without altering the original routing outcome or requiring additional training. The main contributions of this work are as follows:

\begin{itemize}[leftmargin=*]
    \item \textbf{Comprehensive Expert Correlation Analysis:} We conduct a comprehensive analysis of expert selection behavior in representative MoE-based LLMs across several applications, and show that expert activations exhibit strong cross-layer and cross-token correlation during decoding. To the best of our knowledge, this is the first systematic documentation of these cross‑layer and cross‑token correlations.
    \item \textbf{Lightweight Expert Prediction Strategy:} We propose a lightweight dynamic expert prediction strategy that leverages a cross-layer correlation table and a history table to predict future expert requests without modifying the original MoE model or introducing additional training overhead.
    \item \textbf{Prediction-Guided Reconfigurable Hardware:} We design a reconfigurable hardware architecture that supports prediction-guided expert prefetching, flexible expert staging, and pipelined execution, effectively hiding expert loading latency and improving both performance and energy efficiency.
\end{itemize}

Evaluation results on real-world applications show that ST-MoE achieves 85\% expert prediction accuracy and delivers average speedups of 2.5x, 2.2x, and 1.5x, as well as average energy efficiency improvements of 2.5x, 1.8x, and 2.0x, compared to GPU, Adap-Gating, and Pre-gated MoE, respectively.

\section{LLM Inference and Mixture-of-Experts}
\label{subsec: LLM Basics}

Given a user prompt, an LLM first uses a pre-determined tokenizer to convert the prompt into a sequence of input tokens for subsequent inference, as illustrated in Fig.~\ref{fig: figure 1}(a). Typically, LLM inference consists of two key phases: {\it prefill} and {\it decoding}, as shown in Fig.~\ref{fig: figure 1}(b). During the prefill phase (LLM iteration 1), the model processes all input tokens in parallel to construct intermediate states. During the decoding phase, in contrast, output tokens are generated sequentially, and each newly generated token depends on the previously generated tokens~\cite{xu2025wsc,li2025h2}. Within each LLM iteration, at each block in the Transformer decoder, the model uses a Mixture-of-Experts (MoE) architecture comprised of $N$ experts implemented as feed-forward networks. Modern LLMs rely on a large and increasing number of experts, typically ranging from 8 to over 256 experts per layer~\cite{du2022glam,hwang2023tutel,liu2024deepseek,meta2025llama,rajbhandari2022deepspeed,shazeer2017outrageously}. The MoE architecture also contains a gating network that is specified through N vectors of parameters $\mathbf{w}_j$, one vector for each expert $1 \leq j \leq N$. Given an arbitrary token with hidden representation $\mathbf{x}$, the gating network computes for each expert a gating score $z_j = \mathbf{w}^T_j \mathbf{x}$ that is then mapped into a gating probability $s_j$ by applying a softmax transform. The experts corresponding to the top $K$ gating probabilities are then selected to be used with the input $\mathbf{x}$, and their outputs are summed up, using their gating probabilities as weights \cite{dai-etal-2024-deepseekmoe}.

\begin{figure}
    \centering
    \includegraphics[width=\linewidth]{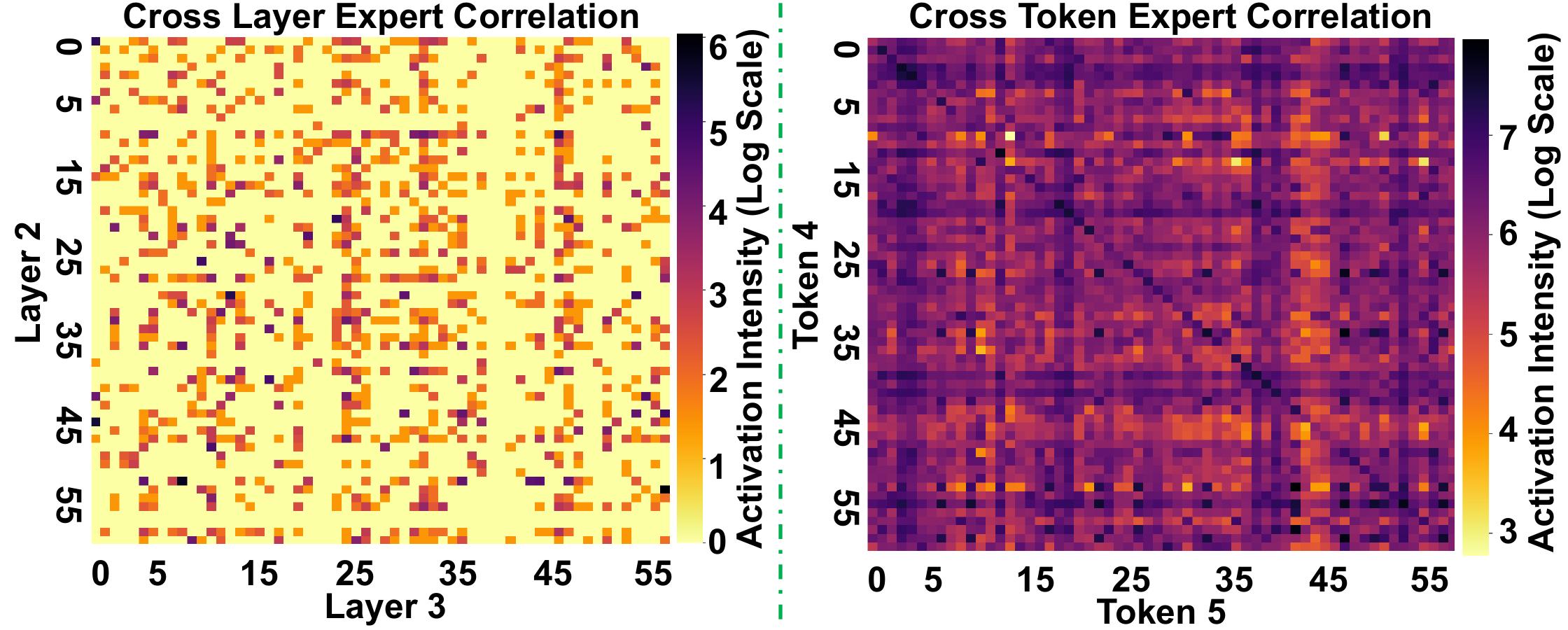}
    \caption{(a) Spatial relationship of different experts across consecutive MoE layers for the same token. (b) Temporal relationship of consecutive decoding tokens within the same MoE layer.}
    \label{fig: correlationship}
\end{figure}

\begin{figure*}
    \centering
    \includegraphics[width=\linewidth]{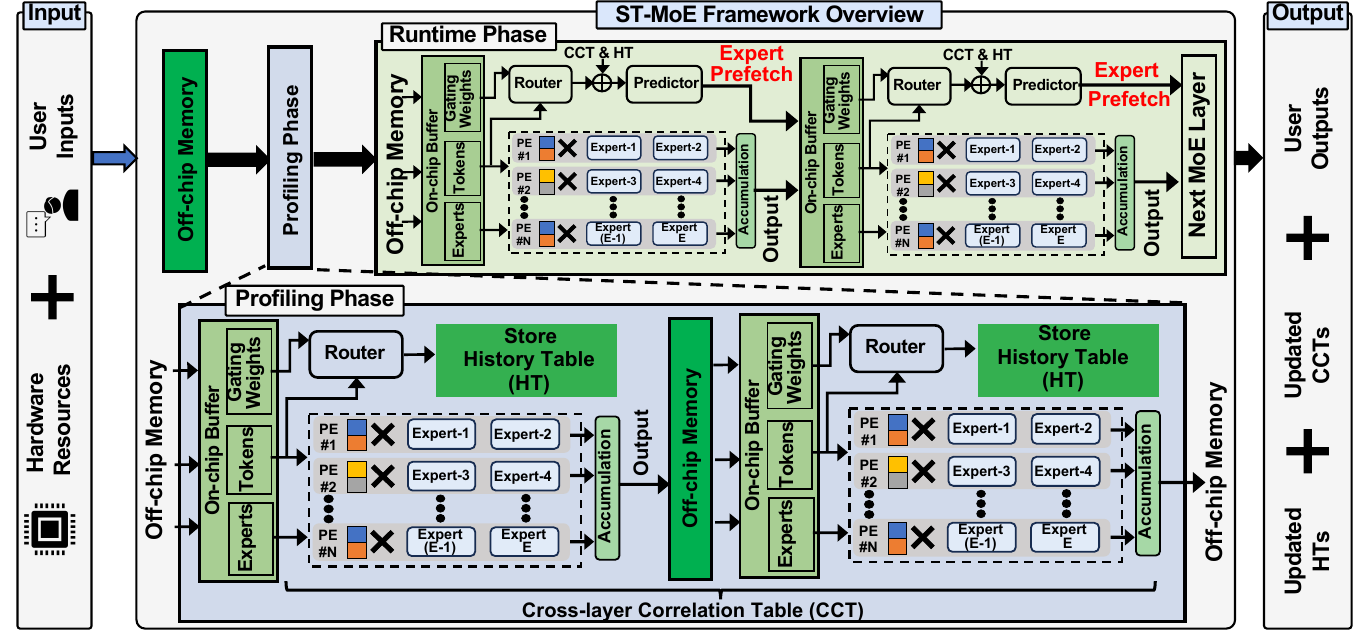}
    \caption{Overview of the proposed ST-MoE framework. The framework consists of two phases: profiling and runtime. The profiling phase constructs the History Table (HT) for each MoE layer and the Cross-layer Correlation Table (CCT), while the runtime phase uses these tables to dynamically predict the experts likely to be activated.}
    \label{fig: framework}
\end{figure*}

\section{Spatio-Temporal Correlations in MoE}
\label{sec: oberstations and motivations}

A key challenge in MoE-based LLM inference is that expert requests are generated dynamically at runtime. However, our analysis shows that, within the same application domain, expert selections are not independent. Instead, they exhibit strong correlation across consecutive layers and between consecutive decoded tokens, which creates an opportunity to predict expert requests and enable effective expert prefetching, improving performance and energy efficiency.

\subsection{Cross-Layer Expert Correlation}
\label{sec:cross-layer}

Upon collecting statistics from multiple MoE-based LLMs, we observe that expert selections for the same token are correlated across consecutive MoE layers. This behavior can be explained by the fact that transformer layers operate on progressively refined representations of the same token while preserving its high-level semantic characteristics, which causes routing decisions in neighboring layers to be statistically dependent. We first analyze the DeepSeek model~\cite{liu2024deepseek,dai2024deepseekmoe} on the CNN Daily Mail dataset \cite{hermann2015teaching,chen2016thorough} using an article summarization workload. Fig.~\ref{fig: correlationship}(a) shows the expert co-activation heatmap between two consecutive MoE layers, Layer 2 and Layer 3, based on 100,000 input tokens. We further perform a chi-squared test on cross layer expert selections, and the resulting \textit{p} values are consistently below 0.01, confirming that expert activations across consecutive layers are not independent. We observe similar strong correlations across different models, including Qwen~\cite{wang2024qwen2,bai2023qwen} and Mixtral~\cite{jiang2024mixtral}, and across different datasets, including HumanEval \cite{chen2021evaluating,peng2024humaneval,cassano2023multipl} and MATH \cite{lightman2023lets,gou2023tora}. 

\subsection{Consecutive-Token Expert Correlation}

Expert selections within the same MoE layer are also observed to be correlated between consecutive decoded tokens. This correlation arises because autoregressive decoding generates each token based on the previously generated context, causing neighboring tokens to exhibit similar semantic dependencies and hidden representations. Consequently, consecutive decoding tokens often trigger overlapping expert selections. We use the same experimental setup as in Section~\ref{sec:cross-layer}, including the models, datasets, and applications. 
Fig.~\ref{fig: correlationship}(b) shows the expert co-activation heatmap for two consecutive decoding tokens within the same MoE layer. We further compare the measured overlap in expert selections between consecutive decoded tokens against a random routing baseline. For an MoE layer with $N$ experts, where each token activates $K$ experts, the expected overlap under independent random routing is E(N) = $K^2/N$. Across multiple LLMs and datasets, the observed overlap consistently exceeds this baseline by nearly 2 times, confirming that expert selections across consecutive decoding tokens are correlated rather than independent.

\section{ST-MoE Framework Design}

Our observations show that expert selections exhibit cross-layer and cross-token correlation during decoding, indicating that future expert requests are predictable from prior routing behavior. 
Motivated by this, we propose an application-aware dynamic expert prefetching framework that combines algorithm and hardware co-design to proactively stage experts before they are needed, thereby reducing expert loading latency and improving energy efficiency without compromising model inference accuracy. In this section, we first present the overview of the proposed framework design in Sec.~\ref{subsec: framework overview}. We then describe the algorithm-level design in Sec.~\ref{subsec: algorithm design}, which exploits both cross-layer and cross-token correlation to predict the experts required by each MoE layer dynamically. Finally, we present the reconfigurable hardware architecture in Sec.~\ref{subsec: hardware design}, which efficiently supports dynamic expert prefetching to improve performance and energy efficiency.

\subsection{Framework Overview}
\label{subsec: framework overview}

Fig.~\ref{fig: framework} illustrates the overall workflow of the proposed ST-MoE framework, which consists of two main phases: \textit{Profiling} and \textit{Runtime}. In the profiling phase, ST-MoE uses a small set of input tokens to capture per-layer expert selection behavior and cross-layer expert correlation. Then, ST-MoE records the resulting information in a History Table (HT) and a Cross layer Correlation Table (CCT), respectively. These tables are used during the runtime phase to guide dynamic expert prediction and prefetching for each MoE layer. During the runtime phase, while the current MoE layer is being executed, ST-MoE uses the gating network's result of the current layer, together with the HT and CCT, to predict the experts likely to be activated in the next MoE layer. The predicted expert indices are then sent to the memory interface of the hardware platform, which proactively prefetches the corresponding experts in parallel with computation, as shown in Fig.~\ref{fig: framework}. For clarity, the attention block is omitted from the figure, while the detailed workflow and the overlap between expert prefetching and computation are described later in Sec.~\ref{subsubsec: pipelined workflow}. When execution reaches the MoE layer, ST-MoE compares the prefetched experts with the results of the gating network, updates the corresponding HT and CCT entries, and retrieves any missing experts caused by misprediction without compromising inference accuracy.

\subsection{Dynamic Prediction Strategy}
\label{subsec: algorithm design}

To predict future activated experts for input tokens in each MoE layer, ST-MoE introduces a dynamic prediction strategy that jointly leverages both spatial and temporal correlation across adjacent MoE layers and consecutive decoding tokens, thereby enabling expert prefetching ahead of execution. To realize this strategy, ST-MoE introduces two key data structures: the Cross-layer Correlation Table (CCT) to capture cross-layer expert correlations and the History Table (HT) to record expert selection history for the previous decoding token. The prediction workflow consists of three steps: table construction, expert prediction, and table update. We first describe the design and initialization of the CCT and HT, and then explain how they are used and updated during execution.

\begin{algorithm}[t]
\caption{Build Cross-layer Correlation Table (CCT)}
\label{alg: build_cct}
\begin{algorithmic}[1]
\STATE \textbf{Input:~} $P$: number of tokens for the profiling phase, $E$: number of experts in each MoE layer, $K$: number of candidate experts per entry.
\STATE \textbf{Output:~}$\mathrm{CCT}[E][2K]$: cross-layer correlation table.

\STATE // 11: strongly preferred, 10: weakly preferred
\STATE // 01: weakly not preferred, 00: strongly not preferred

\STATE \textbf{Initialize:}
\STATE $\mathrm{Co\mbox{-}matrix}[E][E] \leftarrow 0$

\FOR{$s = 1$ to $\mathrm{P}$}
    \STATE $E_i \leftarrow$ experts selected at layer $i$
    \STATE $F_i \leftarrow$ experts selected at layer $(i+1)$
    \FORALL{$e \in E_i$}
        \FORALL{$f \in F_i$}
            \STATE $\mathrm{Co\mbox{-}matrix}[e][f] \leftarrow \mathrm{Co\mbox{-}matrix}[e][f] + 1$
        \ENDFOR
    \ENDFOR
\ENDFOR

\FOR{$e = 1$ to $E$}
    \STATE $\mathrm{CCT}[e][*] \leftarrow$ Top-$K$ experts in $\mathrm{Co\mbox{-}matrix}[e][f]$
    \STATE $\mathrm{CCT}[e][*]_{ConfScore} \leftarrow 10$
\ENDFOR

\STATE \textbf{return} $\mathrm{CCT}$
\end{algorithmic}
\end{algorithm}

\textbf{Cross-layer Correlation Table (CCT):} The CCT is designed to capture correlations among selected experts across adjacent MoE layers, as outlined in Algorithm~\ref{alg: build_cct}. For each expert in the current MoE layer, the CCT stores the most strongly correlated experts in the next MoE layer, providing cross-layer guidance for future expert prediction. To build the CCT, ST-MoE first profiles a small set of input tokens and records expert co-activation patterns for each pair of adjacent layers (Line 7-12). Based on these statistics, it constructs a correlation heatmap for each adjacent layer pair, where each entry indicates how often an expert in the current layer is associated with an expert in the next layer. ST-MoE then selects the $Top-K$ correlated experts in the next layer for each expert in the current layer and stores them in the corresponding CCT entry, where $K$ is set to match the number of activated experts in the target LLM (Line 13-15). In this way, the CCT transforms dense cross-layer correlation statistics into a compact lookup structure that enables efficient expert prediction in the Runtime phase.

As expert selection patterns evolve during inference processing, the CCT must be updated dynamically to maintain accurate cross-layer correlation. Inspired by conventional branch predictor design, ST-MoE associates each CCT entry with a 2-bit confidence score for every stored correlated expert. Accordingly, each score has four possible states: \textit{00} denotes strongly not preferred, \textit{01} denotes not preferred, \textit{10} denotes preferred, and \textit{11} denotes strongly preferred. All entries are initialized to \textit{10}, so each stored expert is initially treated as a preferred candidate (Line 15 in Algorithm~\ref{alg: build_cct}). During execution, the confidence score is updated based on whether the corresponding expert matches the actual result of the gating network. Once the score falls below the preferred region, defined as 0 in our design, the associated expert becomes eligible for replacement by another expert not currently stored in the CCT. The replacement policy is described later as part of the overall workflow.

\textbf{History Table (HT):} The HT captures correlations across consecutive decoding tokens within the same MoE layer. For each MoE layer, the HT stores the actual $Top-K$ experts selected by the gating network for the previous decoding token. As discussed in Sec.~\ref{sec: oberstations and motivations}, consecutive tokens in autoregressive decoding often exhibit correlated routing behavior. Therefore, the expert selections of the previous token provide useful temporal information for predicting the experts required by the current token. Similar to the CCT, each HT stores $K$ experts, where $K$ follows the same $Top$-$K$ routing rule as the target LLM model. Specifically, the HT records the actual experts selected by the gating network for the previous decoding token in each MoE layer. For consistency with the CCT-based prediction logic, each stored expert is also associated with a 2-bit confidence score. Unlike the CCT, however, these scores are not updated during execution and are initialized to \textit{10}. After each token is processed, the HT is overwritten with the latest $Top$-$K$ routing results, so that it always reflects the expert selection outcome of the immediately preceding token. Therefore, the HT provides a lightweight and reliable representation of cross-token correlation.

\begin{algorithm}[t]
\caption{Expert Predict Function}
\label{alg: predict}
\begin{algorithmic}[1]
\STATE \textbf{Input:~}$t$: current token index, $E_i$: experts selected at layer $i$ for token $t$, $\mathrm{CCT}$: cross-layer correlation table, $\mathrm{HT}$: history table
\STATE \textbf{Output:~}$\mathrm{PrefetchSet}$: experts to prefetch for layer $(i+1)$

\STATE \textbf{Initialize} 
\STATE $\mathrm{ConfScore}[E] \leftarrow 0$ for all experts $E$

\FORALL{expert $e \in E_i$}
    \FORALL{candidate expert $f \in \mathrm{CCT}[e]$}
        \STATE $\mathrm{ConfScore}[f] \leftarrow \mathrm{ConfScore}[f] + \mathrm{CCT}[e][f]$
    \ENDFOR
\ENDFOR

\FORALL{expert $h \in \mathrm{HT}$}
    \STATE $\mathrm{ConfScore}[h] \leftarrow 1$
\ENDFOR

\FORALL{expert $h \in \mathrm{HT}$}
    \STATE $\mathrm{ConfScore}[h] \leftarrow \mathrm{ConfScore}[h] + 1$
\ENDFOR

\STATE $\mathrm{PrefetchSet} \leftarrow \{\, f \mid \mathrm{ConfScore}[f] \geq 2 \,\}$
\STATE \textbf{return} $\mathrm{PrefetchSet}$
\end{algorithmic}
\end{algorithm}

\textbf{Prediction Workflow:} Based on the CCT and HT, ST-MoE performs dynamic expert prediction and table maintenance using two functions: Expert Predict and Table Update, as detailed in Algorithm-\ref{alg: predict} and Algorithm-\ref{alg: update}, respectively. In the predict function, once the results of the gating network for the current MoE layer become available, ST-MoE predicts the expert requests for the next MoE layer by jointly using the CCT and HT. Specifically, the selected experts of the current layer are used to index the corresponding CCT entries, producing a set of spatially correlated candidate experts for the next layer (Line 5-7). Subsequently, ST-MoE accesses the HT entry of the next layer to obtain temporally correlated expert candidates derived from the previous decoding token (Line 10-11). The candidate experts from the two sources are merged, and their confidence scores are aggregated. If an expert appears in both the CCT and HT, its final confidence score is computed by combining the two scores, as shown in Eq.~\ref{eq: confidence score}; otherwise, its score remains unchanged. Finally, ST-MoE selects all experts whose confidence scores are no smaller than the threshold as the predicted expert set for the next MoE layer and forwards their indices to the memory interface of the hardware platform for proactive prefetching. In our prediction algorithm, this threshold is set to 2, corresponding to the 2-bit confidence state (10).

\begin{equation}
\begin{aligned}
\mathrm{Score}(e)=
\begin{cases}
\mathrm{CCT}(e)+\mathrm{HT}(e), & \text{if } e \in \mathrm{CCT} \text{ and } e \in \mathrm{HT},\\
\mathrm{CCT}(e), & \text{if } e \in \mathrm{CCT} \text{ and } e \notin \mathrm{HT},\\
\mathrm{HT}(e), & \text{if } e \notin \mathrm{CCT} \text{ and } e \in \mathrm{HT}.
\end{cases}
\end{aligned}
\label{eq: confidence score}
\end{equation}

For the Table Update function, as shown in Algorithm-\ref{alg: update}, once the actual result of the gating network for the predicted MoE layer becomes available, ST-MoE compares it with the predicted expert set and updates the corresponding CCT and HT entries accordingly. For the CCT, if a predicted expert is confirmed by the actual routing result, its confidence score is increased by 1 (Line 7), with the maximum value capped at \textit{11}, which denotes the strongly preferred state. Otherwise, its confidence score is decreased by 1 (Line 10). If the confidence score falls below \textit{0} in our design, an expert replacement procedure is triggered. In this case, the corresponding expert is replaced with a new candidate expert (Line 12-14), and the confidence score of the new expert is initialized to \textit{10}, indicating the preferred state. Meanwhile, the HT is overwritten with the latest $Top$-$K$ expert selection results so that it always reflects the routing outcome of the immediately preceding token. Through this online update process, ST-MoE continuously adapts the CCT and HT to evolving expert selection patterns, thereby improving the prediction accuracy of subsequent MoE layers.

\begin{algorithm}[t]
\caption{Table Update Function}
\label{alg: update}
\begin{algorithmic}[1]
\STATE \textbf{Input: }$E_i$: experts selected at layer $i$, $F_j$: experts selected at layer $(i+1)$, $\mathrm{CCT}$: cross-layer correlation table
\STATE \textbf{Output: }Updated $\mathrm{CCT}$

\FORALL{expert $e \in E_i$}
    \STATE $\mathrm{PrefetchSet} \leftarrow$ experts stored in $\mathrm{CCT}[e][*]$
    \FORALL{$(f,\mathrm{ConfScore}) \in \mathrm{CCT}[e][*]$}
        \IF{$f \in F_j$}
            \STATE $\mathrm{ConfScore} \leftarrow \min(\mathrm{ConfScore}+1,\,3)$
        \ELSE
            \IF{$\mathrm{ConfScore} > 0$}
                \STATE $\mathrm{ConfScore} \leftarrow \mathrm{ConfScore}-1$
            \ELSE
                \STATE $g \leftarrow$ any expert in $(F_j \setminus \mathrm{PrefetchSet})$
                \STATE $f \leftarrow g$
                \STATE $\mathrm{ConfScore} \leftarrow 10$
            \ENDIF
        \ENDIF
    \ENDFOR
\ENDFOR

\STATE \textbf{return} $\mathrm{CCT}$
\end{algorithmic}
\end{algorithm}

\subsection{Reconfigurable Architecture Design}
\label{subsec: hardware design}

To enable the proposed dynamic expert prediction strategy in MoE inference, we implement a reconfigurable architecture that integrates expert prefetching support, flexible compute resources, and an adaptive interconnection network. In ST-MoE, reconfigurability is exposed through three coordinated mechanisms: expert-to-processing element (PE) mapping, permutation-network-based data routing, and PE-level dataflow configuration. These mechanisms are driven by prediction-guided expert availability, where the predicted expert set determines which expert weights are proactively staged on chip, while the actual gating results determine how the staged experts are mapped, routed, and executed. By coupling expert prefetching with runtime mapping, configurable data routing, and PE-level dataflow adaptation, the architecture overlaps expert transfer with computation while maintaining energy efficiency. In the following, we first introduce an overview of the architecture, followed by the prefetching hardware, the reconfigurable compute engine and interconnection design, and the overall pipelined execution workflow.

\subsubsection{\textbf{Architecture Overview:}} 
\begin{figure}
    \centering
    \includegraphics[width=\linewidth]{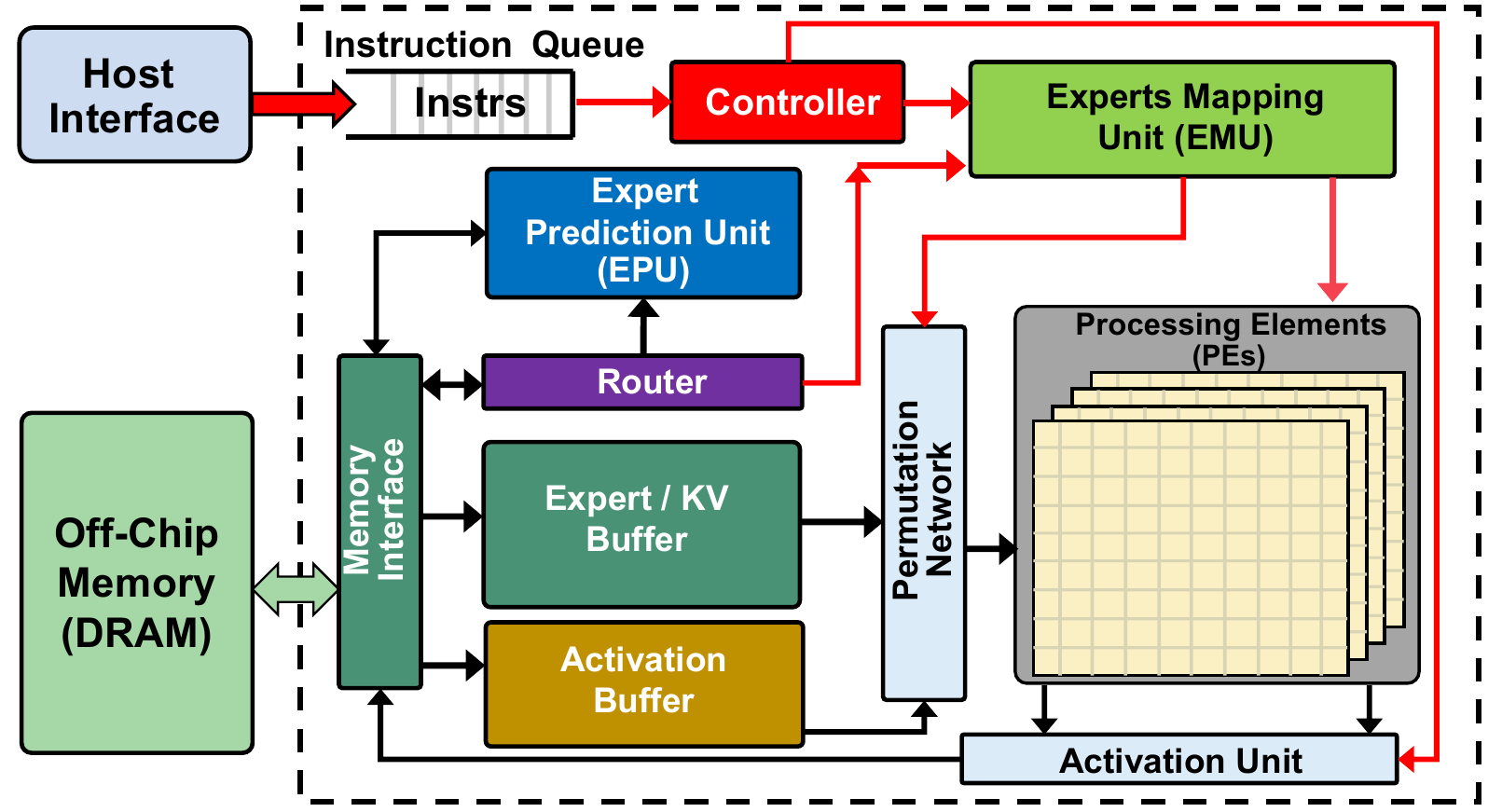}
    \caption{Overview of the proposed ST-MoE hardware platform, which consists of four key components: the Expert Prediction Unit (EPU) for algorithm-level prediction strategy, the router for gating network, the unified computing engine for key matrix computations, and the Expert Mapping Unit (EMU) for workload distribution.}
    \label{fig: arch overview}
\end{figure}

Fig.~\ref{fig: arch overview} illustrates the overall architecture of the proposed ST-MoE hardware platform, which is designed to support prediction-guided expert prefetching for efficient MoE-based LLM inference. The architecture consists of multiple key hardware components, including the Expert Prediction Unit (EPU), Expert Mapping Unit (EMU), Router, Expert and KV Buffer, Activation Buffer, Processing Elements (PEs), Permutation Network, and Activation Unit. Specifically, during execution, the router performs the gating network and generates the actual expert selection pattern for the current input token. The EPU uses this routing information, together with the CCT and HT, to predict the experts likely to be required by the subsequent MoE layer. The predicted expert indices are then sent to the memory interface, which proactively fetches the corresponding expert weights from off-chip memory and stages them in the Expert/KV Buffer. When the actual gating result of a MoE layer becomes available, the EMU dynamically maps the selected experts to available PEs and configures the permutation network to route the selected expert weights and token activations from the on-chip buffers to the assigned PEs. The PEs, permutation network, and activation unit form the reconfigurable compute engine, where each PE can configure its local buffer usage and dataflow according to the reuse opportunity of the current expert computation.

\subsubsection{\textbf{Prefetching Hardware:}} To support the proposed dynamic expert prefetching strategy, ST-MoE uses the Router, the Expert Prediction Unit (EPU), the Expert Mapping Unit (EMU), the memory interface, and on-chip buffers. The router includes an $M \times K$ MAC array to execute the matrix multiplication in the gating network and generate the expert selection patterns for the input tokens, where $M$ denotes the input batch size and $K$ denotes the number of selected experts under the $Top$-$K$ routing strategy of the target LLM. The EPU uses a small local buffer to store the CCT and HT, and performs lightweight confidence-score computation based on the current routing result to predict the experts likely to be selected in the next MoE layer. The EPU also supports runtime updates to the prediction-related tables, allowing the prediction behavior to adapt to evolving expert activation patterns. The predicted expert indices are forwarded to the memory interface, which proactively fetches the corresponding expert weights from off-chip memory and stages them in the Expert/KV Buffer. In this way, expert prefetching is driven by runtime prediction rather than by a static schedule. After the actual gating result becomes available, the EMU translates the dynamic expert selection into an executable mapping. Specifically, the EMU assigns the selected experts to available PEs based on the current expert requests and the availability of prefetched experts in the on-chip buffer. It then configures the permutation network to route the selected expert weights and token activations to the assigned PEs. This mapping and routing process provides runtime reconfiguration for irregular MoE expert activation patterns. Together, these components enable efficient coordination of expert prediction, expert staging, expert-to-PE mapping, and data delivery during MoE execution.

\begin{figure}
    \centering
    \includegraphics[width=\linewidth]{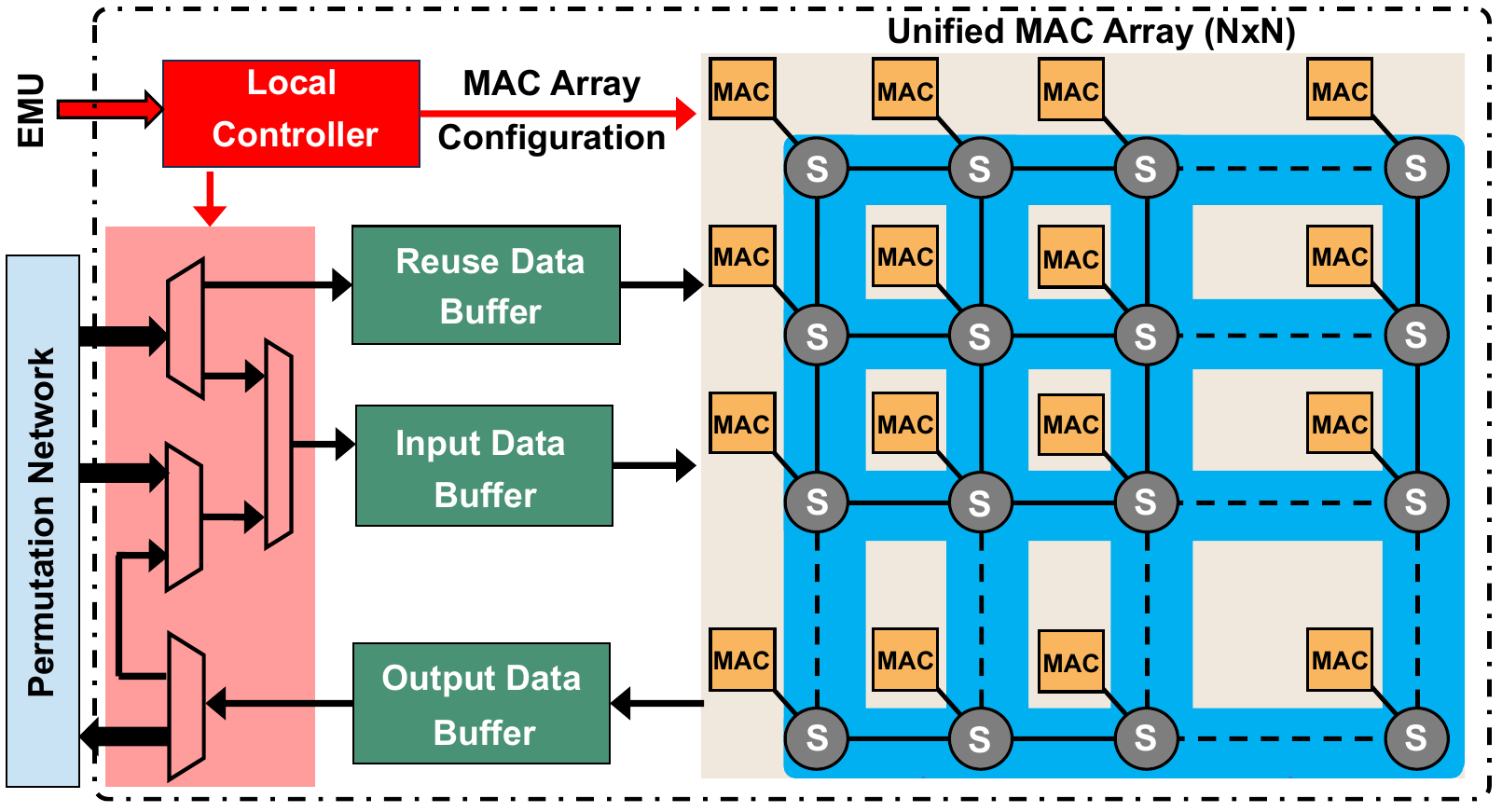}
    \caption{Architecture overview of the Processing Element (PE). It includes two input buffers for input token-related data and weight data, respectively, an NxN MAC array for matrix-matrix multiplication, and a dedicated output buffer for storing intermediate and final results. The MUX and DeMUX units are controlled by the local controller to manage local data movement.}
    \label{fig: PE design}
\end{figure}

\subsubsection{\textbf{Computation Hardware:}} ST-MoE implements a central compute array composed of $N$ Processing Elements (PEs) to execute the main computations of each MoE layer, such as matrix multiplication. Fig.~\ref{fig: PE design} shows the detailed architecture of a PE, which includes a local controller, a unified MAC array, a reuse data buffer, an input data buffer, an output data buffer, and MUX and DeMUX units. Each PE serves as a reconfigurable compute unit whose local buffer usage, MUX/DeMUX paths, and dataflow can be configured according to the workload assigned by the EMU.

After receiving the workload assignment from the EMU, the local controller configures the PE according to the dimensions of the input activations and the expert weight matrix. The controller first determines the most suitable dataflow for the current computation, such as input stationary or weight stationary, depending on which operand offers higher reuse opportunity under the assigned workload. Based on the selected dataflow, the controller configures the reuse data buffer and input data buffer so that the appropriate data can remain locally in the MAC array and be reused across multiple computation cycles. This PE-level reconfiguration does not rely on a new standalone systolic dataflow; instead, it allows the compute engine to adapt conventional dataflows to the dynamic expert assignments and prefetched expert availability produced by ST-MoE.

The local controller then configures the internal switches of the unified MAC array to match the selected dataflow. To efficiently support the dense matrix multiplication required by each MoE layer, the MAC units are organized in a systolic manner, allowing partial sums and reused operands to propagate through neighboring MAC units in a regular pattern. Such an organization improves throughput and reduces fine-grained control overhead, making it well-suited for the repeated matrix computations involved in MoE layers. After computation, the generated intermediate or final outputs are stored in the output data buffer. When these outputs are required by the subsequent layer without additional off-chip transfer, the PE can directly forward the buffered results back to the input data buffer through the local MUX and DeMUX units. This mechanism allows the PE to reuse locally produced data in the next round of computation, thus reducing unnecessary buffer-to-memory traffic and shortening the data delivery path between consecutive computations.

\subsubsection{\textbf{Pipelined Workflow:}~}
\label{subsubsec: pipelined workflow}
\begin{figure}
    \centering
    \includegraphics[width=\linewidth]{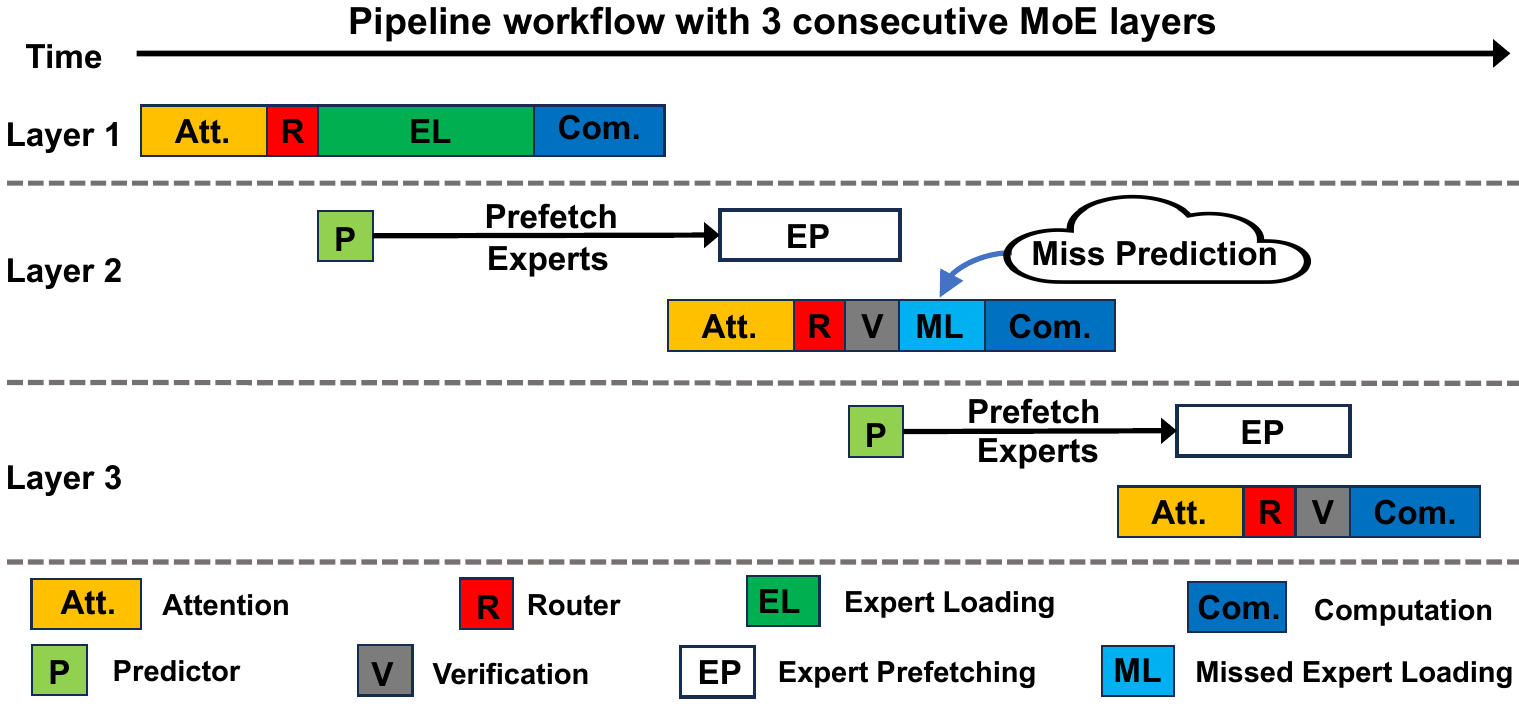}
    \caption{Pipeline workflow of three consecutive MoE layers. \textit{Att.} is the attention operation, \textit{R} is gating network, \textit{EL} is expert loading, \textit{Com.} is expert computation, \textit{P} predicts the experts required by the subsequent layer, \textit{V} performs verification and table update, \textit{EP} is expert prefetching, and \textit{ML} indicates missed loading due to expert misprediction. The figure illustrates how ST-MoE overlaps expert prefetching with attention computation while handling mispredicted experts through verification.}
    \label{fig: pipelined workflow}
\end{figure}

ST-MoE organizes MoE inference as a pipelined workflow that overlaps expert prediction, prefetching, execution, and verification across adjacent MoE layers. Typically, each MoE layer consists of four main stages: attention, gating, expert loading, and expert computation. For ST-MoE, rather than waiting until the next layer requires expert loading, it initiates prediction and prefetching early so that expert transfer can proceed in parallel with ongoing computation. To illustrate the pipelined execution flow, we use three consecutive MoE layers as an example, as shown in Fig.~\ref{fig: pipelined workflow}. After receiving the result from the gating network, Layer 1 proceeds to expert loading, followed by expert computation. Simultaneously, ST-MoE uses the gating result of Layer 1 together with the CCT and HT loaded from the off-chip memory to predict the experts likely to be required by the following Layer 2. Once the prediction is completed, the predicted expert indices are temporarily buffered in the EPU, since Layer 2 must still go through its attention stage before expert prefetching is issued. When Layer 2 approaches the end of attention, the predicted expert set is sent from the EPU buffer to the prefetching hardware, which transfers the corresponding experts from off-chip memory into the on-chip buffers before Layer 2 begins expert computation. In this way, expert transfer for Layer 2 is overlapped with the computation of the attention block.

When execution advances to Layer 2, its attention and gating stages are first completed as usual, while the required experts are still being prefetched. Once the gating result of Layer 2 becomes available, ST-MoE verifies the set of prefetched experts against the actual selected experts. If some experts required by Layer 2 are not included in the prefetched set due to misprediction, as shown in Fig.~\ref{fig: pipelined workflow}, the missing experts are fetched immediately to preserve inference correctness. At the same time, the verification outcome of Layer 2 is used to update the HT with the latest $Top-K$ gating result and to adjust the corresponding confidence scores in the CCT. Meanwhile, the gating result of Layer 2 is reused by the EPU to trigger the next round of expert prediction for Layer 3. The same procedure is then repeated across subsequent layers, thereby forming a continuous pipelined workflow for expert prediction, transfer, execution, and update. By overlapping expert prediction and prefetching with the computation of the current layer, ST-MoE hides a substantial portion of expert loading latency. Meanwhile, the online verification and table updates ensure that prediction errors do not affect inference correctness, allowing ST-MoE to improve efficiency without compromising inference accuracy.

\section{Evaluation and Analysis}

\subsection{Evaluation Setup}

\begin{table}[]
\renewcommand\arraystretch{1.3}
\centering
\caption{Detailed MoE-based LLM models}
\label{tab:dataset}
\begin{scriptsize}
\setlength{\tabcolsep}{3.5mm}
\begin{tabular}{|c|c|c|c|c|} 
\hline
\rowcolor[HTML]{C0C0C0} 
\textbf{Series} & \textbf{Model} & \textbf{\#Layers} & \textbf{\#Experts} & \textbf{\#Top-K} \\ \hline
\multirow{2}{*}{Qwen} & Qwen 1.5 & 24 & 60 & Top-4 \\ \cline{2-5} 
 & Qwen 2.0 & 28 & 64 & Top-6 \\ \hline
\multirow{2}{*}{DeepSeek} & DeepSeek V2 & 60 & 160 & Top-8 \\ \cline{2-5} 
 & DeepSeek MoE & 60 & 128 & Top-8 \\ \hline
\end{tabular}
\end{scriptsize}
\end{table}

\begin{table}[]
\begin{footnotesize}
\centering
\renewcommand{\arraystretch}{1.2}
\caption{Dataset statistics for profiling and evaluation (P and E denotes profiling and evaluation)}
\label{tab:dataset_stats}
\setlength{\tabcolsep}{3.5pt}
\begin{tabular}{|c|c|c|c|}
\hline
\rowcolor[HTML]{C0C0C0} 
Dataset & Task & Samples (P / E) & \# Decoded Tokens \\
\hline
CNN/DM    & Article Summarization & 35,000 / 145,000  &  \textasciitilde 19.8 M \\
\hline
MATH      & Math Reasoning & 1,500 / 6,500 &  \textasciitilde 1.2 M \\
\hline
HumanEval & Code Generation  & 30 / 134 &  \textasciitilde 0.05 M \\
\hline
\end{tabular}
\end{footnotesize}
\end{table}

\textbf{Evaluation Design.~}We extend SCALE-Sim~\cite{samajdar2018scale} with a cycle-accurate hardware simulator implemented in C++ to model the execution behavior of the proposed ST-MoE framework. To obtain execution time, the simulator counts the number of on-chip memory access and the arithmetic operations performed by each PE. The number of on-chip memory access is used to estimate the on-chip communication time, while the number of arithmetic operations is used to calculate the computation time. The off-chip communication time is evaluated using DRAMsim2~\cite{rosenfeld2011dramsim2}. The overall execution time is evaluated through a comprehensive analysis of temporal overlaps across different memory hierarchies, integrating the durations of on-chip communication, off-chip communication, and computational phases. During data transfer, the system leverages buffering techniques to enable concurrent activities, such as simultaneously loading the next tile from off-chip memory while distributing the current on-chip tile to PEs and executing computational tasks. The simulator models these parallel activities by determining the maximum duration across communication and computation stages, thereby providing a precise result. Additionally, the simulator calculates the energy consumption by multiplying the total power consumption with the total execution time, providing a comprehensive assessment of the architecture's energetic performance. For area consumption, we implement all proposed hardware-related logic through the Synopsys Design Compiler with the TSMC 40nm library for the synthesis. We set the clock frequency at 1 GHz. We use Cacti 6.0~\cite{muralimanohar2009cacti} to estimate the area, power, and access latency of all types of on-chip buffers.

\textbf{Hardware Configuration.~}For the hardware platform, we implement the proposed ST-MoE with \textit{K} processing elements (PEs), where each PE serves as an independent compute unit for a specific selected expert during each MoE layer. In our evaluation, we set \textit{K = 8}. Each PE integrates a (64 $\times$ 64) MAC array. All MAC units are interconnected into a systolic manner to effectively perform matrix multiplication, resulting in a total of 32,768 MAC units throughout the system. In addition, we conduct an ablation study in Section~\ref{subsubsec: ablation study} to evaluate the impact of varying the size of the MAC array within each PE on overall performance. The routing unit includes a (512 $\times$ 8) MAC array to execute the gating network, where 512 denotes the input batch size of the LLM model, and 8 corresponds to the number of PEs in the hardware platform. All MAC units use bfloat16 (BF16) arithmetic.

\begin{table}
    \centering
    \begin{small}
    \caption{Key hardware components of ST-MoE and PE}
    \begin{tabular}{|c|c|c|c|}
    \hline
    \rowcolor[HTML]{C0C0C0} 
    \textbf{ST-MoE} & \textbf{Value} & \textbf{Area ($mm^2)$} & \textbf{Power (W)} \\ \hline
    Processing Element Array & 1x8 & 426.1 & 50.6 \\ \hline
    Expert/KV Buffer & 16 MB & 131.1 & 4.3 \\ \hline
    Activation Buffer & 4 MB & 32.8 & 1.1 \\ \hline
    Expert Prediction Unit & - & 0.1 &  0.02\\ \hline
    Router & - & 28.7 & 5.5 \\ \hline
    DRAM & 256 GB/s & - & - \\ \hline
    \hline
    \rowcolor[HTML]{EFEFEF} 
    \textbf{PE} & \textbf{Value} & \textbf{Area ($mm^2)$} & \textbf{Power (mW)} \\ \hline
    Reuse Data Buffer & 2 MB &  16.4 & 0.5 \\ \hline
    Input \& Output Buffer & 512 KB & 4.1 & 0.1 \\ \hline
    Unified MAC Array & 64 x 64& 28.6 & 5.8 \\ \hline    
    \end{tabular}
    \label{tab: key hardware components}
\end{small}
\end{table} 

\noindent\textbf{Benchmarks.~}We evaluate the proposed ST-MoE using two representative set of MoE-based LLMs, Qwen and DeepSeek, which differ in model scale and expert routing characteristics. As summarized in Table-\ref{tab:dataset}, these models vary in the number of layers, the number of experts per MoE layer, and the \textit{Top-K} routing policy, providing a diverse evaluation space for analyzing ST-MoE under different expert activation patterns and model configurations. To cover representative real-world workloads, we use applications from article summarization, math reasoning, and code generation, including CNN Daily Mail (CNN/DM), MATH, and HumanEval. For each workload, a part of input samples is used for the profiling stage to construct the CCT, while the remaining samples are used to evaluate the prediction accuracy of the proposed dynamic expert prediction strategy, the total execution time, and the total energy consumption, as shown in Table-\ref{tab:dataset_stats}. Token counts are estimated using the Qwen1.5 tokenizer as a unified reference.

\begin{figure}
    \centering
    \includegraphics[width=\linewidth]{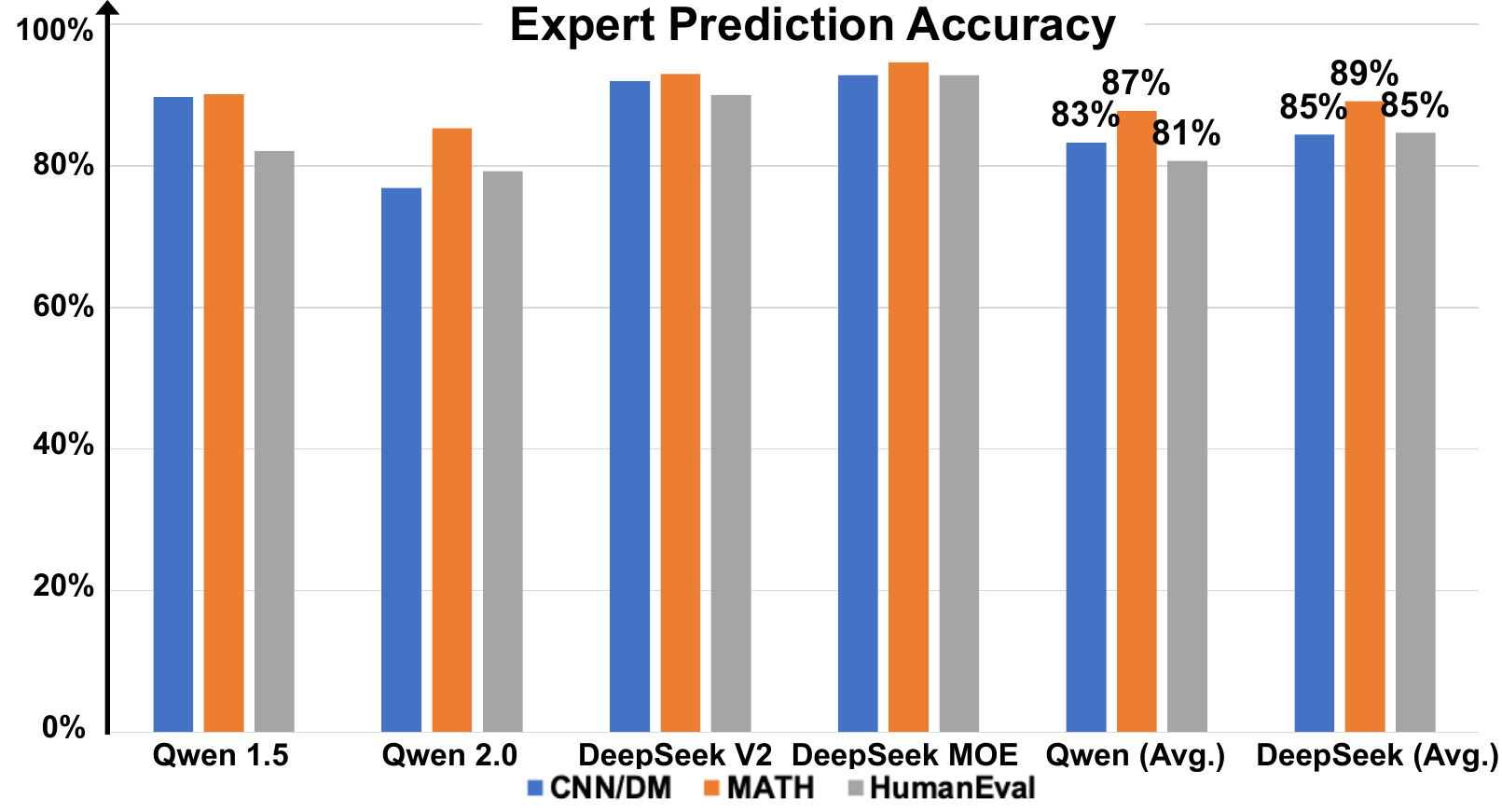}
    \caption{Prediction accuracy of the proposed dynamic expert prefetching strategy across different LLMs and applications.}
    \label{fig: prediction accuracy}
\end{figure}

\noindent\textbf{Baselines.~}To perform cross-platform comparisons, we evaluate the proposed ST-MoE framework against both a conventional GPU platform and prior customized hardware designs. Specifically, we use the NVIDIA A100 GPU as the general-purpose baseline, denoted as PyGT-GPU. Additionally, we compare the proposed ST-MoE with two recent representative hardware accelerators for MoE-based LLM inference, namely Adap-gating and Pre-gated MoE, denoted as Adap-G and Pre-gated, respectively. To ensure a fair comparison, we normalize the hardware resources of all customized accelerators to match the proposed ST-MoE architecture in terms of total MAC count and off-chip memory bandwidth. In this way, the compared designs operate under the same compute and memory resource budget. For the GPU baseline, we report the performance obtained from the original platform configuration, since its architecture is fixed and cannot be scaled in the same manner as customized accelerators. This evaluation methodology ensures that the observed differences among customized designs primarily reflect architectural efficiency, rather than the advantages of raw hardware capacity. 

\begin{figure*}
    \centering
    \includegraphics[width=\linewidth]{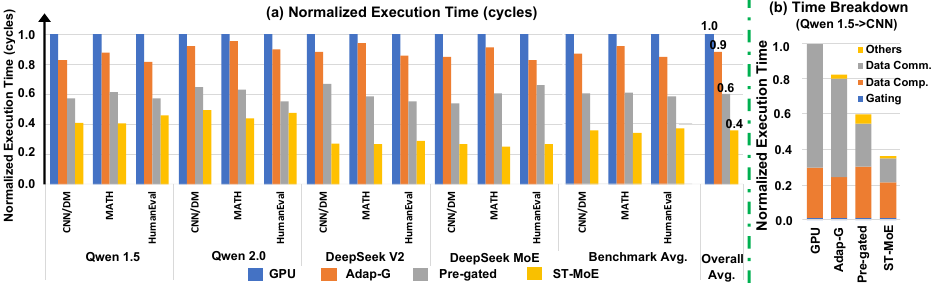}
    \caption{(a) Normalized execution time (cycles) of ST-MoE compared to state-of-the-art baselines across diverse LLMs and real-world applications. (b) Normalized execution time breakdown for Qwen1.5 on CNN/DM in the article summarization task. All results are normalized to the GPU baseline, and lower is better.}
    \label{fig: execution time}
\end{figure*}

\subsection{Area and Power Analysis~}

To evaluate the area and power consumption of the proposed ST-MoE hardware design, we use Synopsys Design Compiler with the TSMC 40nm standard cell library. Table~\ref{tab: key hardware components} summarizes the area and power consumption of the key hardware components in the proposed ST-MoE and PEs. Overall, the hardware cost is dominated by the compute and storage components, including the PE array, the expert/KV buffer, and the activation buffer. In particular, the PE array accounts for the largest share of both area (66\%) and power consumption (81\%), while the expert/KV buffer also contributes a noticeable area cost due to the storage capacity required for expert staging and KV data buffering. In contrast, the hardware overhead introduced by the prediction and prefetching support is minimal. Specifically, in our design, the EPU stores a 256-entry CCT, where each entry contains 8 candidates of 10 bits, including 8 bits for the expert candidate index and 2 bits for the confidence score. In addition, the HT contains 8 entries, each with 10 bits. Together with a small number of additional MAC units for confidence score accumulation during prediction, the proposed EPU incurs only about 0.02\% area overhead, while its power overhead is nearly negligible relative to the total power consumption. These results indicate that the proposed dynamic expert prediction strategy can be implemented with minimal hardware overhead, making ST-MoE a practical design for improving the inference efficiency of MoE-based LLMs.

\begin{figure*}
    \centering
    \includegraphics[width=\linewidth]{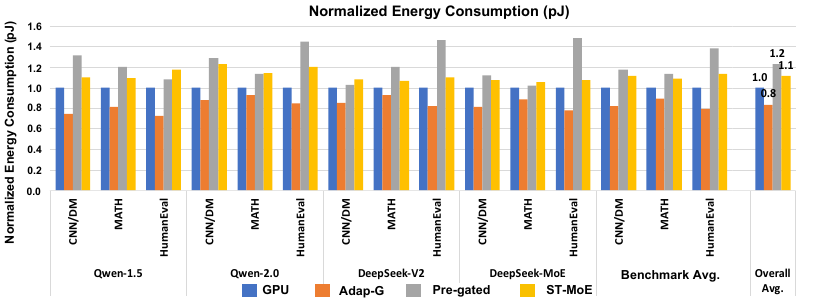}
    \caption{Normalized energy consumption (pJ) of ST-MoE compared to state-of-the-art baselines across diverse LLMs and real-world applications. The energy overhead of ST-MoE is mainly caused by the miss penalty of the expert prediction strategy. All results are normalized to the GPU baseline, and lower values indicate better energy efficiency.}
    \label{fig: energy consumption}
\end{figure*}

\subsection{Expert Prediction Accuracy} 

Fig.~\ref{fig: prediction accuracy} shows the accuracy of the proposed dynamic expert prediction strategy for different MoE-based LLMs and real-world applications. In general, ST-MoE achieves high prediction accuracy on all evaluated benchmarks, with most cases exceeding 80\%, demonstrating that future expert requests can be effectively predicted using the proposed lightweight table-based strategy. A clear trend is that MATH consistently achieves higher prediction accuracy than the other applications. This is because mathematical reasoning typically follows more structured and constrained decoding patterns, leading to more stable token-level dependencies and more consistent expert activation behavior across both consecutive layers and decoding steps. In contrast, applications such as general article summarization on CNN/DM dataset often exhibit more diverse token transitions and less regular routing behavior, making expert prediction relatively more challenging. Despite these variations, ST-MoE maintains a sufficiently high accuracy across all cases, providing a strong foundation for proactive expert prefetching and for overlapping expert loading with computation.

\subsection{End-to-End Execution Time:~}

Fig.~\ref{fig: execution time}(a) shows the normalized end-to-end execution time (cycles) of ST-MoE and the baseline designs across different MoE-based LLMs and real-world applications. Overall, ST-MoE consistently achieves the lowest execution time in all evaluated benchmarks, demonstrating the effectiveness of the proposed prediction-guided expert prefetching framework. On average, ST-MoE reduces the overall execution time by 60\%, 56\%, and 33\% compared to GPU, Adap-gating, and Pre-gated MoE, respectively. By predicting future expert requests and prefetching the corresponding experts before they are needed, ST-MoE overlaps expert transfer with ongoing computation, thereby reducing the effective latency of off-chip expert memory access.

\begin{figure*}
    \centering
    \includegraphics[width=\linewidth]{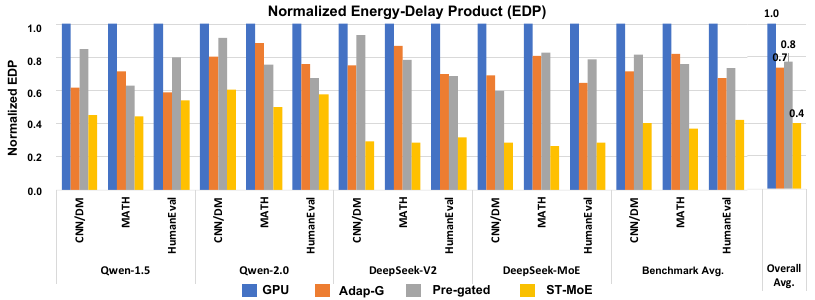}
    \caption{Normalized energy-delay product (EDP) of ST-MoE versus state-of-the-art baselines across diverse LLMs and real-world applications. All results are normalized to the GPU baseline. (lower is better)}
    \label{fig: EDP}
\end{figure*}

The large gap between ST-MoE and the conventional GPU baseline highlights the cost of dynamic expert loading under a general-purpose execution model. On conventional GPUs, expert weights are fetched on demand for each MoE layer, which incurs substantial data movement. As a result, expert access remains a major bottleneck and leads to the longest execution time among all LLM models. Although Adap-G improves efficiency by dynamically reducing the number of activated experts in each MoE layer instead of using a fixed number, its benefit depends on suppressing expert usage. ST-MoE instead retains the full routing outcome and reduces execution time through prediction-guided prefetching, avoiding the efficiency accuracy trade-off introduced by adaptive expert suppression. Compared to Pre-gated MoE, the advantage of ST-MoE lies in how future expert requests are predicted. Pre-gated MoE depends on an additional trained pre-gating function, whereas ST-MoE directly leverages the cross-layer and cross-token correlation to make the prediction. This allows ST-MoE to achieve lower execution time without additional training overhead or deployment complexity. Although ST-MoE still has the potential to mispredict experts, the resulting miss penalty is bounded by the high prediction accuracy shown in Fig.~\ref{fig: prediction accuracy}. Fig.~\ref{fig: execution time}(b) also shows a time breakdown for Qwen 1.5 on the CNN/DM dataset, illustrating the contribution of each component, including the gating network, computation, and data communication, to the overall execution time reduction. Clearly, the majority of the time reduction comes from lower data communication overhead, particularly reduced expert transfer.

\subsection{End-to-End Energy Consumption:~}Fig.~\ref{fig: energy consumption} shows the normalized energy consumption (pJ) of ST-MoE and the baseline designs across different MoE-based LLMs and real-world applications. Compared to the conventional GPU baseline, ST-MoE incurs a modest average energy overhead of 10\%. This overhead is primarily caused by prediction mismatches. When the prefetched experts do not fully match the actual routing result after performing the gating network of the current MoE layer, ST-MoE issues additional off-chip memory access to fetch those missing experts, which increases total energy consumption while preserving inference accuracy. In contrast, Adap-G consumes less energy than the conventional GPU baseline because it dynamically lowers the effective $Top-K$ value and thus the number of activated experts, and the associated memory access and energy consumption. Pre-gated MoE also incurs additional energy overhead because it relies on an extra pre-gating function and proactive expert transfer, both of which introduce extra computation and data movement beyond conventional MoE execution.

\textbf{Energy-Delay Product (EDP): }To better capture overall energy efficiency, we further evaluate the Energy-Delay Product (EDP), which jointly accounts for execution time and energy consumption, as shown in Fig.~\ref{fig: EDP}. Because ST-MoE substantially reduces execution time while introducing only a modest energy overhead, EDP serves as a more comprehensive metric for evaluating the practical efficiency of the proposed framework. Across different LLM models and real-world applications, ST-MoE improves EDP by 2.5x, 1.8x, and 2.0x, compared to GPU, Adap-G, and Pre-gated MoE, respectively.

\begin{figure}
    \centering
    \includegraphics[width=\linewidth]{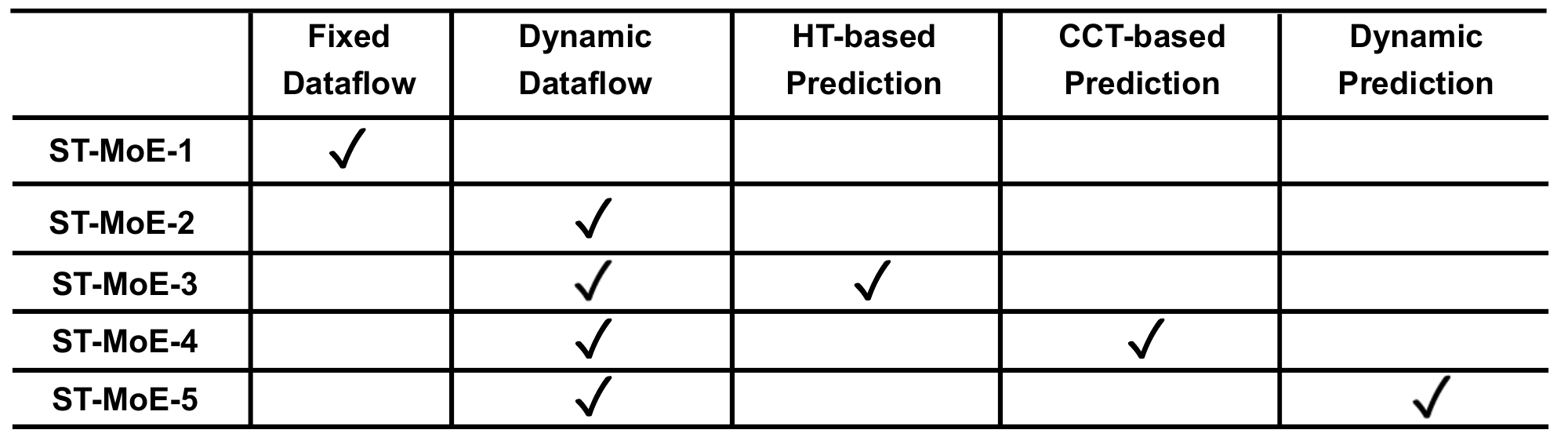}
    \caption{Five different ST-MoE configurations for the ablation study.}
    \label{fig: ablation table}
\end{figure}

\subsection{Ablation Study}
\label{subsubsec: ablation study}

\begin{figure}
    \centering
    \includegraphics[width=\linewidth]{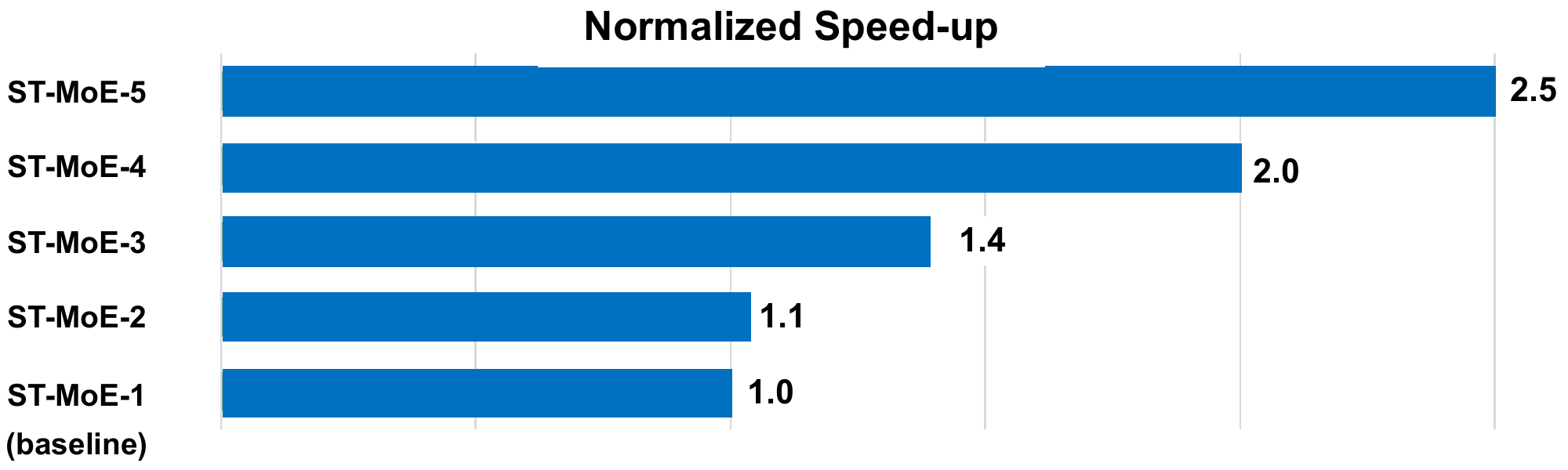}
    \caption{An ablation study of ST-MoE with five distinct configurations. All values are normalized to ST-MoE-1.}
    \label{fig: ablation study}
\end{figure}

To quantitatively evaluate the contribution of each proposed design component, we conduct an ablation study using Qwen across all evaluated applications. Fig.~\ref{fig: ablation table} details the five ST-MoE configurations considered in this study, while Fig.~\ref{fig: ablation study} presents the average normalized speedup, with all results normalized to the hardware-only baseline using the weight stationary (WS) dataflow. Starting from this baseline, enabling dynamic dataflow improves performance by allowing the compute hardware to adapt to different workload characteristics and improve local data reuse efficiency. Adding temporal correlation provides additional improvement, showing that expert selection information from previous decoding tokens complements the cross-layer prediction. Finally, incorporating spatial correlation further improves performance by enabling cross-layer expert prediction and proactive expert prefetching. By combining the hardware support, joint spatial-temporal prediction, and dynamic dataflow selection, the full ST-MoE design achieves the highest speedup among all evaluated configurations.

\subsection{Hardware Sensitivity Analysis}

\begin{figure}
    \centering
    \includegraphics[width=\linewidth]{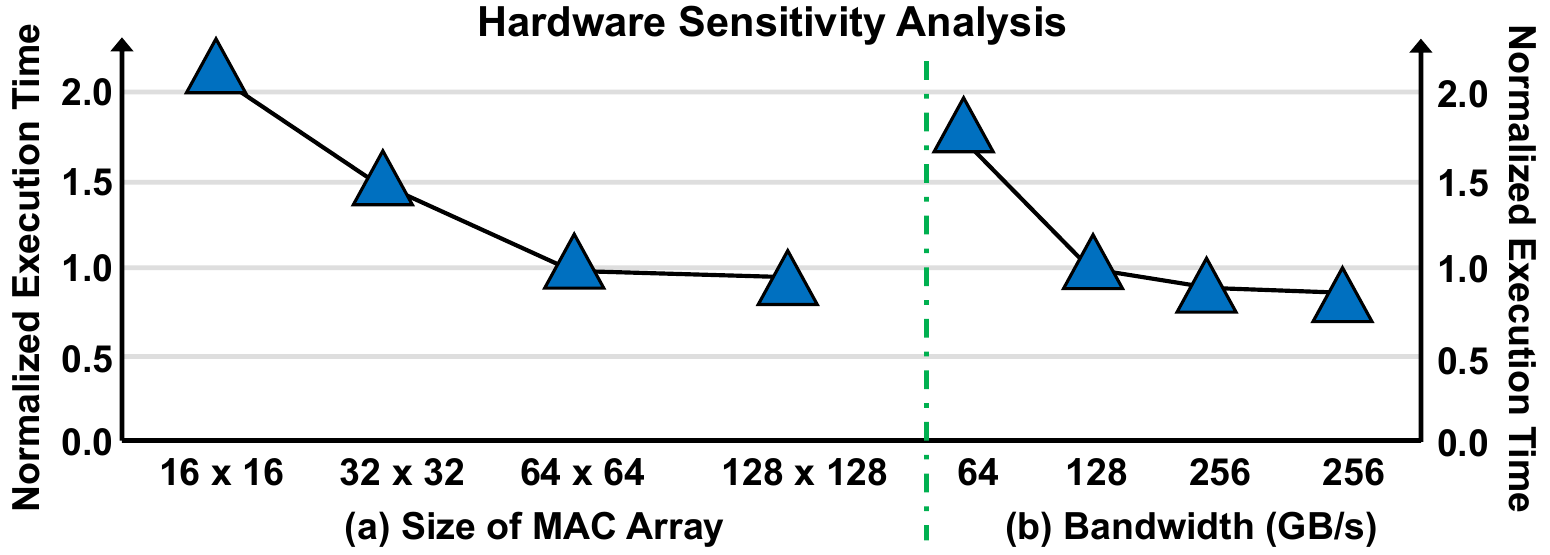}
    \caption{Hardware sensitivity analysis of ST-MoE: (a) normalized execution time under varying MAC array sizes within each PE; (b) normalized execution time under varying memory bandwidths.}
    \label{fig: Hardware sensitivity analysis}
\end{figure}

To evaluate the impact of key hardware resources on ST-MoE, we conduct a hardware sensitivity analysis by varying the MAC array size within each PE and the off-chip memory bandwidth, which directly affect compute capability and expert transfer efficiency, respectively. Fig.~\ref{fig: Hardware sensitivity analysis} shows the normalized execution time under different hardware configurations. All results are normalized to the design with a 64$\times$64 MAC array and 256 GB/s off-chip memory bandwidth. As the MAC array size increases, each PE can process more token computations in parallel, which reduces execution time. However, the performance gain gradually saturates at larger MAC array sizes, as the communication overhead among MAC units becomes increasingly significant. Similarly, increasing off-chip memory bandwidth reduces execution time by accelerating expert transfer, thereby improving the overlap between expert loading and computation. This benefit, however, also gradually diminishes as expert transfer ceases to be the dominant bottleneck. Overall, these results show that ST-MoE scales effectively with both compute capability and memory bandwidth, while delivering the greatest benefit.

\section{Related Work}

\textbf{Efficient MoE-based LLM Inference.~}Recent work has explored several directions for improving the inference efficiency of MoE-based LLMs. Specifically, Adap-gating~\cite{li2023adaptive} reduces end-to-end inference latency by dynamically adjusting the number of activated experts in each MoE layer. However, this approach modifies the routing behavior of the original model and therefore affects model accuracy. Pre-gated MoE~\cite{hwang2024pre} modifies the role of the gate function by using the gate in the current MoE block to preemptively select the experts for the next MoE block, thereby moving the expert-selection dependency across adjacent blocks and enabling proactive expert transfer. However, this approach requires an additional training procedure for the pre-gated function, and changes how the original gating mechanism is used during inference. Similarly, SiDA~\cite{du2024sida} and PreScope~\cite{yu2025prescope} rely on additional trained predictors to improve expert prediction quality, but introduce extra training overhead and limited portability. MoE-Infinity~\cite{xue2024moe} exploits skewed expert reuse patterns during decoding to guide expert caching. However, its optimization is centered on retaining frequently reused experts in the cache, whereas ST-MoE proactively stages future experts by modeling correlations across layers and decoding tokens. Fate~\cite{fang2025fate} proposes a cross-layer gate mechanism that clones the current gate input to the CPU and runs the next-layer gate in parallel to generate expert prefetching decisions. By relying on CPU-side gate execution and CPU-GPU coordination, Fate is primarily suitable for CPU-GPU offloading systems. Overall, prior work improves the efficiency of MoE-based LLMs by modifying routing behavior, training auxiliary predictors, expert caching, or CPU-side gate execution. In contrast, the ST-MoE prefetching framework described in this paper jointly exploits cross-layer and consecutive-token expert activation correlations, together with a lightweight prediction strategy, to enable training-free and gate-preserving expert prefetching with reconfigurable architecture support.

\textbf{LLM Accelerator Design.~}A large body of prior work has proposed customized hardware architectures for accelerating Transformer and LLM inference through optimized compute engines, memory hierarchies, and interconnection fabrics~\cite{li2025h2,wei2025spatial,chen2025bitmod,lee2025paise,xu2025wsc}. Specifically, $H^2$LLM~\cite{li2025h2} proposes a heterogeneous accelerator based on hybrid-bonding for low-batch LLM inference and performs hardware–dataflow co-exploration to balance computational capacity and bandwidth while optimizing both prefill and decoding execution. The works in~\cite{wei2025spatial,xu2025wsc} adopt wafer-scale hardware designs with flexible interconnection fabrics to efficiently support LLM inference. BitMoD~\cite{chen2025bitmod} introduces fine-grained data type adaptation, which assigns different numerical data types to groups of weights to enable very low precision quantization while preserving accuracy. To support this flexibility, it employs bit-serial processing elements that efficiently handle multiple precisions and data types, although this design may introduce additional adaptation and control overhead. Despite their effectiveness, none of these customized LLM accelerators is specifically tailored for MoE layer execution during LLM inference, although MoE layers contribute a large portion of the overall execution time. Consequently, they do not directly address the core challenges of dynamically predicting, staging, and executing experts, which are critical to efficient MoE-based LLM inference.

\textbf{Matrix-matrix Multiplication Design.~}Since the major computation pattern in each MoE layer is matrix-matrix multiplication, another relevant line of research is hardware acceleration for matrix multiplication, including sparse-dense matrix multiplication (SpMM), general matrix multiplication (GeMM), and sparse matrix–vector multiplication (SpMV)~\cite{hegde2019extensor,pal2018outerspace,qin2020sigma,srivastava2020matraptor,srivastava2020tensaurus}. Extensor~\cite{hegde2019extensor} is designed to accelerate sparse tensor algebra. OuterSPACE~\cite{pal2018outerspace} and MatRaptor~\cite{srivastava2020matraptor} explore different loop unrolling techniques to improve SpMM efficiency. SIGMA~\cite{qin2020sigma} and Tensaurus~\cite{srivastava2020tensaurus} also focus primarily on accelerating SpMM operations. However, most state-of-the-art matrix multiplication accelerators are optimized for sparse workloads, whereas MoE layers are dominated by dense matrix multiplication. Directly applying these prior designs to MoE inference can therefore lead to inefficient execution, as they do not fully exploit the data parallelism of dense matrix computation. Although conventional Tensor Processing Units (TPUs) are specifically designed for dense matrix multiplication, they still leave two important challenges unexplored for MoE inference. First, they lack dedicated hardware support to leverage spatial and temporal correlation for dynamic expert prediction and prefetching. Second, they do not explicitly exploit expert reuse patterns across different input tokens, which may result in frequent access to the same experts and cause substantial redundant memory traffic.

Overall, to the best of our knowledge, ST-MoE is the first expert prefetching framework that implements a lightweight prediction strategy based on spatial cross-layer correlations and temporal consecutive-token correlations among active experts. By using the predictions to proactively stage experts on chip and execute them through reconfigurable hardware support, ST-MoE hides expert memory access latency and significantly improves both performance and energy efficiency compared to prior work.

\section{Conclusions}

In this paper, we analyze expert selection behavior in MoE-based LLMs across diverse real-world applications and observe that expert activations exhibit strong spatial correlation across adjacent layers and temporal correlation across consecutive decoding tokens. Based on this observation, we propose ST-MoE, a framework that jointly leverages spatial and temporal correlations in expert activations to enable prediction-guided expert prefetching, thereby overlapping expert loading with computation. ST-MoE combines a lightweight dynamic expert prediction strategy with a reconfigurable hardware platform to improve performance and energy efficiency without compromising inference accuracy. Experimental results show that ST-MoE achieves 85\% expert prediction accuracy and delivers average speedups of 2.5x, 2.2x, and 1.5x, along with average energy efficiency improvements of 2.5x, 1.8x, and 2.0x, compared to GPU, Adap-Gating, and Pre-gated MoE, respectively.

\bibliographystyle{ACM-Reference-Format}
\bibliography{refs}

\end{document}